# Estimators for Long Range Dependence: An Empirical Study

**William Rea**

*e-mail:* `bill.rea@cantebury.ac.nz`

and

**Les Oxley**

*e-mail:* `les.oxley@cantebury.ac.nz`

and

**Marco Reale**

*e-mail:* `marco.reale@cantebury.ac.nz`

and

**Jennifer Brown**

*e-mail:* `jennifer.brown@cantebury.ac.nz`

**Abstract:**
We present the results of a simulation study into the properties of 12 different estimators of the Hurst parameter, $H$, or the fractional integration parameter, $d$, in long memory time series. We compare and contrast their performance on simulated Fractional Gaussian Noises and fractionally integrated series with lengths between 100 and 10,000 data points and $H$ values between 0.55 and 0.90 or $d$ values between 0.05 and 0.40. We apply all 12 estimators to the Campito Mountain data and estimate the accuracy of their estimates using the Beran goodness of fit test for long memory time series.

MCS code: 37M10

**Keywords and phrases:** Strong dependence, Global dependence, Long range dependence, Hurst parameter estimators.

## 1. Introduction

The subject of long-memory time series was brought to prominence by Hurst (1951) and has subsequently received extensive attention in the literature. See the volumes by Beran (1994), Embrechts and Maejima (2002), and Palma (2007) and the collections of Doukhan et al. (2003) and Robinson (2003) and the references therein.

Of critical importance in analyzing and modeling long memory time series is estimating the strength of the long-range dependence. Two measures are commonly used. The parameter $H$, known as the Hurst or self-similarity parameter,

*

was introduced to applied statistics by Mandelbrot and van Ness (1968) and arises naturally from the study of self-similar processes. The other measure, the fractional integration parameter, $d$, arises from the generalization of the Box-Jenkins ARIMA(p,d,q) models from integer to non-integer values of the integration parameter $d$. This generalization was accomplished independently by Granger and Joyeux (1980) and by Hosking (1981). The fractional integration parameter $d$ is also the discrete time counterpart to the self-similarity parameter $H$ and the two are related by the simple formula $H = d + 1/2$.

A number of estimators of $H$ and $d$ have been developed. These are usually validated by an appeal to some aspect of self-similarity, or by an asymptotic analysis of the distributional properties of the estimator as the length of the time series converges to infinity.

A number of theoretical results on the asymptotic properties of various estimators have been obtained. The aggregated variance method was shown to be asymptotically biased of the order $1/\log N$, where $N$ is the number of observations by Giraitis et al. (1999) who also showed the GPH (Geweke and Porter-Hudak, 1983) estimator was asymptotically normal and unbiased. Robinson (1994) proved the averaged periodogram estimator was consistent under very mild conditions. Lobato and Robinson (1996) obtained its limiting distribution. The Peng et al. (1994) estimator was proved to be asymptotically unbiased by Taqqu et al. (1995). Some theoretical properties of the R/S estimator have been examined by Mandelbrot (1975) and Mandelbrot and Taqqu (1979). Mandelbrot (1975) proved that the R/S statistic is robust to the increment process having a long-tailed distribution in the sense that $E[X_i^2] = \infty$. However, Bhattacharya et al. (1983) proved that the R/S statistic was not robust to departures from stationarity. Thus for a short memory process with slowly decaying deterministic trend the R/S statistic will report an estimate of $H$ which implies the presence of long-memory. An estimator based on wavelets was proved asymptotically unbiased and efficient by Abry et al. (1998). They also showed the traditional variance type estimators were fundamentally flawed and could not lead to good estimators of $H$. Fox and Taqqu (1986) proved the Whittle estimator was consistent and asymptotically normal for Gaussian long range dependent sequences. Dalhaus (1989) proved the estimator of Fox and Taqqu (1986) was efficient. Further theoretical results on the Whittle estimator can be found in Horvath and Shao (1999).

Because the finite sample properties of these estimators can be quite different from their asymptotic properties some previous authors have undertaken empirical comparisons of estimators of $H$ and $d$. Nine estimators were discussed in some detail by Taqqu et al. (1995) who carried out an empirical study of these estimators for a single series length of 10,000 data points, five values of both $H$ and $d$, and 50 replications. Teverovsky and Taqqu (1999) showed in a simulation study that the differenced variance estimator was unbiased for five values of $H$ (0.5, 0.6, 0.7, 0.8, and 0.9) for series with 10,000 observations whereas the aggregated variance estimator was downwards biased. Jensen (1999) undertook a comparison of two estimators based on wavelets, one proposed by Jensen (1999) and the other proposed by McCoy and Walden (1996), with the GPH estimator for four series lengths ($2^7$, $2^8$, $2^9$, $2^{10}$ observations), five values of $d$ and 1000 replications. They reported the wavelet estimators had lower mean squared errors (MSEs) than the GPH estimator for all $d$ values and series lengths investigated. Jeong et al. (2007) carried out a comparison of six estimators on simulated fractional Gaussian noises (FGNs) with 32,768 ($2^{15}$) observations, five values of $H$ and 100 replications.

Several of the above empirical investigations would have been limited by the then available computer power which has since increased considerably. We have extended these studies to a larger number of parameters, higher number of replications and 12 estimators as detailed in Section (2) below.

The remainder of the paper is organized as follows. Section (2) gives details of the method. Section (3) presents the results. Section (4) applies the methods to the Campito Mountain data which is regarded as a standard example of a long memory time series. Section (5) contains the discussion and Section (6) gives our conclusions and suggests avenues of future research.

## 2. Method

Ten $H$ estimators are implemented in the contributed package `fSeries` of Wuertz (2005) for the popular statistical software R (R Development Core Team, 2005). They are the absolute value, aggregated variance, boxed periodogram, differenced variance, Higuchi, Peng, periodogram, rescaled range, wavelet, and the Whittle. The wavelet estimator is discussed in some detail by Abry and Veitch (1998), Abry et al. (1998), and Veitch and Abry (1999) and the other nine are discussed by Taqqu et al. (1995). Further, the GPH (Geweke and Porter-Hudak, 1983) and Haslett and Raftery (1989) are implemented as estimators for $d$ in the contributed package `fracdiff` of Fraley et al. (2006).

Taqqu et al. (1995) simulated FGNs and the corresponding discrete time fractionally integrated (FI(d)) series and found that each estimator performed similarly whether estimating $H$ in simulated FGNs or $d$ in simulated FI(d)s. For example, if an estimator was biased when estimating $H$ it was also biased in a very similar manner when estimating $d$. Thus, with the exception of the GPH and Haslett-Raftery estimators, we only investigated each estimator's performance in estimating $H$ for simulated FGNs. FGNs were generated using the function `fgnSim` in `fSeries`. We ran 1000 replications of simulated FGNs with 100 different lengths and eight different $H$ values. The lengths were between 100 and 10,000 data points in steps of 100. The $H$ values were between 0.55 and 0.90 in steps of 0.05. For each series $H$ was estimated by each of these ten estimators. For each $H$ value and series length we estimated the median, 75% and 95% confidence intervals empirically from the simulated data. The $H$ or $d$ estimates were sorted into ascending order and the median obtained by averaging the 500th and 501st values. Similar calculations were done for the upper and lower values of the 75% and 95% confidence intervals.

For the GPH and Haslett-Raftery estimators we generated FI(d) series with the function `farimaSim` in `fSeries` over the range 0.05 to 0.40 in steps of 0.05.

The other details are the same as above. In the presentation of the results we converted the GPH and Haslett-Raftery $d$ estimates to $H$ equivalents to facilitate comparisons among the estimators.

The simulations and estimations were performed on a SunBlade 1000 with a 750Mhz UltraSPARC-III CPU with 2Gb of memory and a Sun Ultra 10 with a 440Mhz UltraSPARC-IIi CPU and 1Gb of memory.

## 3. Results

To present the results in tabular form would require a very large amount of space. Thus we present them in graphical form. Figures (1) through (7) present some of the results. Figures (1) through (6) are presented with the vertical axis with a range of 1.2 $H$ units to facilitate comparisons among the estimators' standard deviation of their estimates. It should be noted that stationary long memory occurs in the range $0.5 < H < 1.0$. Baillie (1996) states that for $1.0 \leq H < 1.5$ the series are non-stationary but mean reverting while for $0 \leq H \leq 0.5$ the series are anti-persistent. Figure (7) presents the mean squared error (MSE) as a function of series length. We report MSE for series lengths greater than or equal to 500 data points. Again the vertical axes all have the same range to facilitate comparisons.

The results for the absolute value of the variance method are presented in Figures (1) (a) and (c). The absolute value of the variance method was unbiased at all series lengths when $H$ was low (0.55 or 0.60) but became progressively biased and underestimated $H$ as $H$ increased.

The results for the aggregated variance method are presented in Figures (1) (b) and (d). The aggregated variance method exhibited bias and underestimated $H$ in short series when $H$ was low. As $H$ increased the estimator became increasingly biased at all series lengths examined. With $H = 0.90$ the true value of $H$ lay above the upper 95% empirical confidence interval for all but the shortest series lengths.

The results for the boxed periodogram method are presented in Figures (2) (a) and (c). The boxed periodogram method was developed specifically to deal with perceived problems with the periodogram estimator. Comparing the boxed periodogram with the unmodified periodogram method in Figures (4) (a) and (c) we can see that for FGNs where the series were short and $H$ was high that the periodogram method was biased towards over estimating $H$. The boxed periodogram was biased towards underestimating $H$ for almost all values of $H$ and series lengths examined.

The results for the differenced variance method are presented in Figures (2) (b) and (d). The differenced variance method had one of largest confidence intervals of the estimators when the series were short but this slowly decreased as sample size increased. Only the GPH, periodogram and wavelet methods had a similarly wide confidence interval for short series. The differenced variance estimator exhibited bias towards over estimating $H$ for any series with less than 7,000 observations. The bias was very serious in the short series. For series longer than about 9,000 observations the estimator exhibited a small amount of bias towards underestimating $H$.

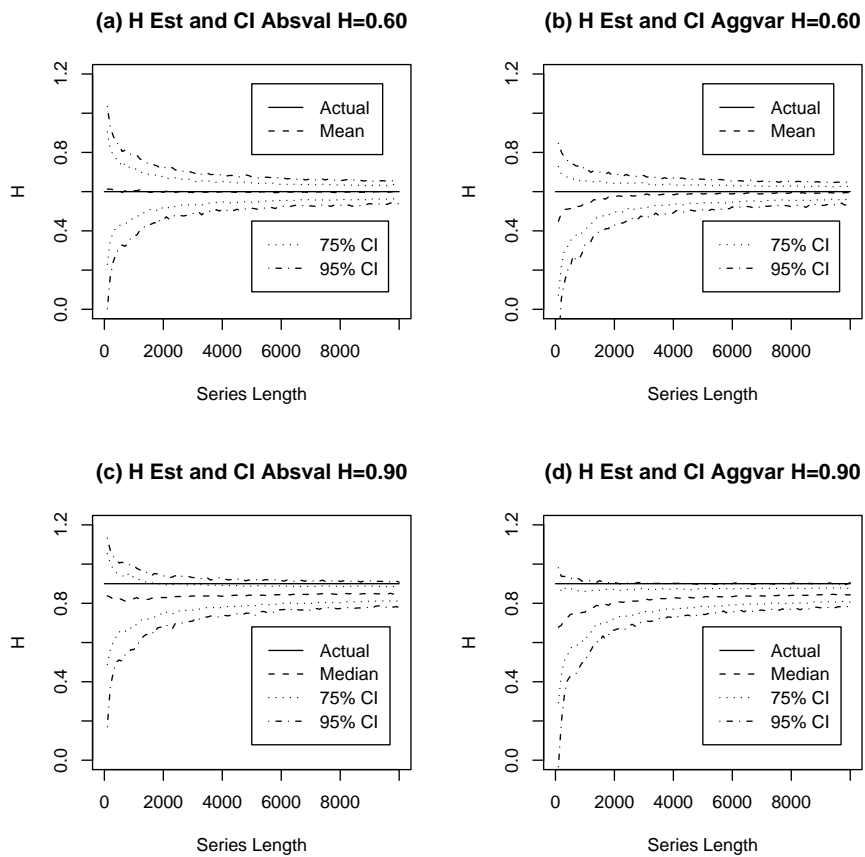

FIG 1. *Empirical confidence intervals for the H estimates with H = 0.60 and H = 0.90; (a) and (c) absolute value method, (b) and (d) aggregated variance estimator.*

The results for the Higuchi (1988) estimator are presented in Figures (3) (a) and (c). The Higuchi was biased towards underestimating $H$ but the magnitude of the bias appeared relatively independent of $H$. The width of the confidence interval of the estimate increased with increasing $H$.

The results for the Peng et al. (1994) estimator are presented in Figures (3) (b) and (d). The Peng estimator was biased toward under estimating $H$ in the series lengths we investigated. This bias appeared to be independent of $H$ but was very small though it appeared greater in short series.

The results for the periodogram estimator were discussed above in conjunction with the boxed periodogram estimator.

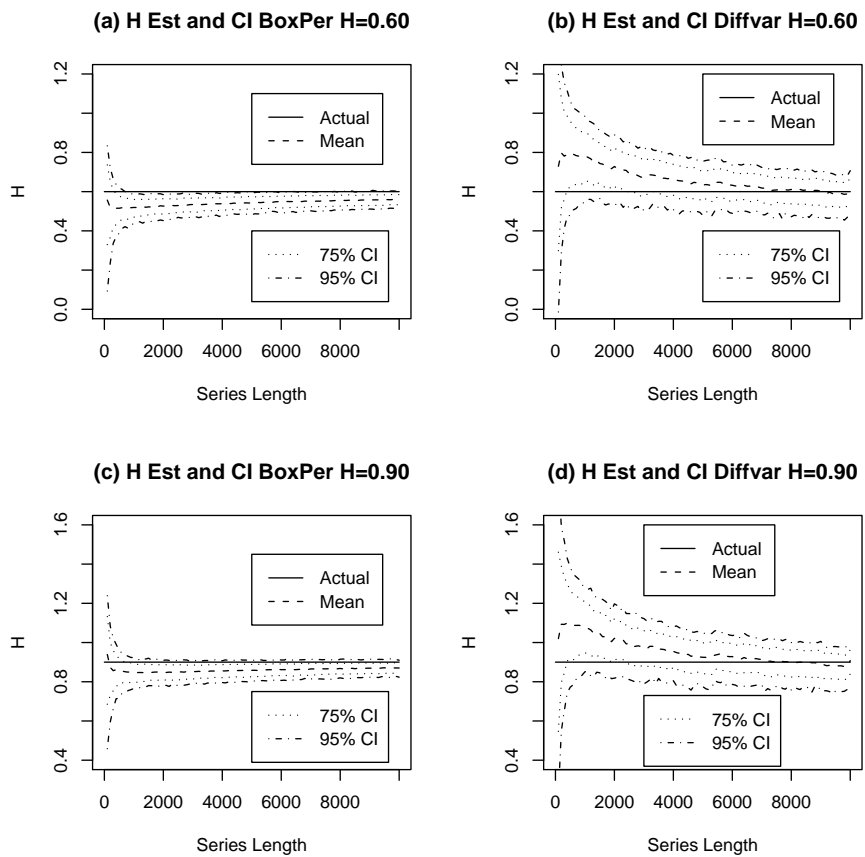

FIG 2. *Empirical confidence intervals for the H estimates with $H = 0.60$ and $H = 0.90$; (a) and (c) boxed periodogram method, (b) and (d) differenced variance estimator.*

The results for the R/S estimator are presented in Figures (4) (b) and (d). The R/S estimator is of considerable historical interest because it was first proposed by Hurst and was used extensively in early studies of long-memory processes. However, as can be seen from Figures (4) (b) and (d) the R/S estimator exhibited three problems; it was biased upwards when $H$ was low, it was biased downwards when $H$ was high, and the confidence interval of the estimate did not decrease with increasing series length once the series reached about 1000 observations.

The results for the Whittle estimator are presented in Figures (5) (a) and (c). Compared to the other nine estimators implemented in `fSeries` the Whittle estimator was remarkable for its narrow confidence interval. It only displayed a small amount of downwards bias when the series were short and $H$ was high. There was an implementation issue in the software we used. The Whittle esti-mator would terminate with an error when $H$ was low and the series contained only a few hundred observations. Thus in Figure (5)(a) there was no data for series with less than 300 observations in the $H = 0.65$ results.

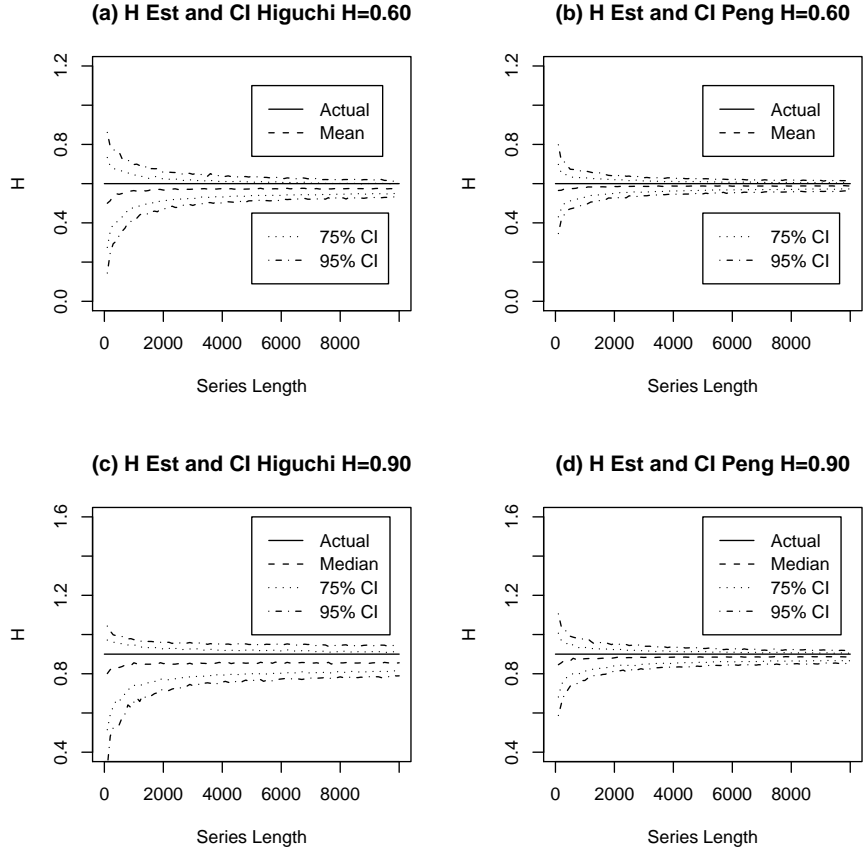

FIG 3. *Empirical confidence intervals for the H estimates with $H = 0.60$ and $H = 0.90$; (a) and (c) Higuchi estimator, (b) and (d) Peng estimator.*

The results for the wavelet estimator are presented in Figures (5) (b) and (d). The wavelet estimator was unbiased for all $H$ values at series lengths over 4,100 data points. The bias present in series shorter than 4,100 data points was very small. The availability of a new *octave* can be seen in Figures (5) (b) and (d) with each doubling of the series length. New *octaves* resulted in a series of steps in the reduction of the confidence interval of the estimate with increasing series length. The estimator had constant variance when the number of *octaves* was fixed.

The results for the GPH estimator are presented in Figures (6)(a) and (c).

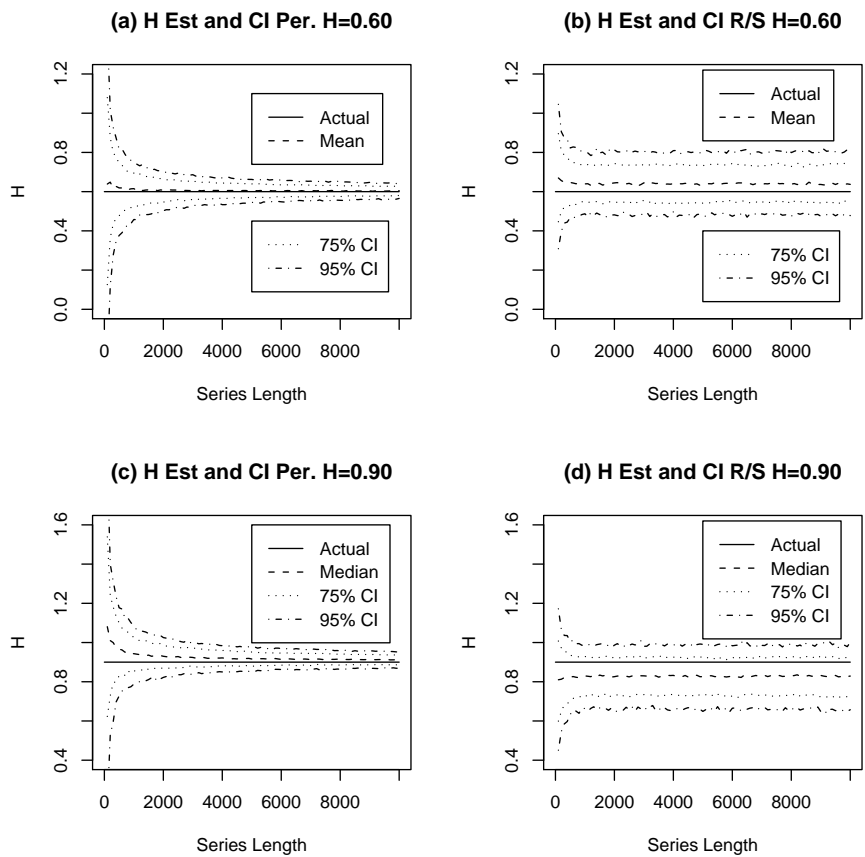

FIG 4. *Empirical confidence intervals for the H estimates with $H = 0.60$ and $H = 0.90$; (a) and (c) periodogram estimator, (b) and (d) R/S estimator.*

The GPH estimator exhibited a very small amount of bias towards overestimating $d$ at all series lengths examined. It had a very wide confidence interval which narrowed slowly as the series length increased.

The results for the Haslett-Raftery estimator are presented in Figures (6)(b) and (d). The Haslett-Raftery did not report estimates of $d$ less than zero ($H < 0.5$). Hence for low $d$ and short series the distribution was truncated on the low side at $d = 0$ or $H = 0.5$ as in Figure (6) (a). The Haslett-Raftery estimator was an excellent estimator with only small amounts of bias in the short series and had a narrow confidence interval.

Figure (7) presents the MSEs for the estimators for $H = 0.9$ or $d = 0.4$ as appropriate. This is an alternative way to look at the data from the simulations. We only report MSEs for series of 500 data points and longer because of the high

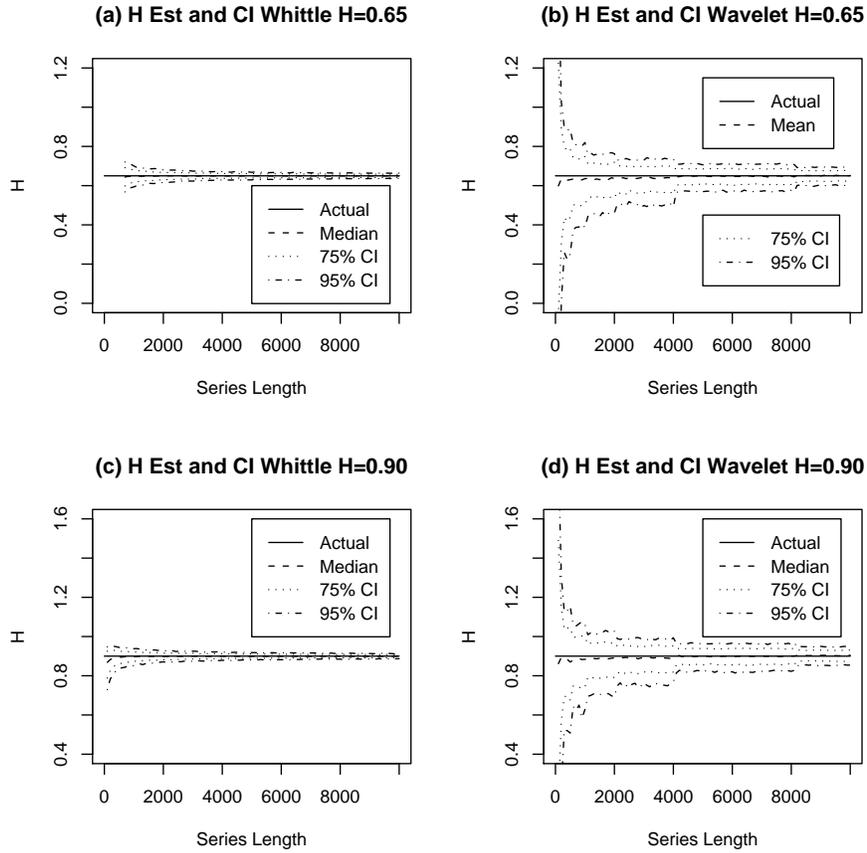

FIG 5. *Empirical confidence intervals for the H estimates with $H = 0.65$ and $H = 0.90$; (a) and (c) Whittle estimator, (b) and (d) wavelet estimator.*

MSEs for some estimators in the short series. The Whittle and Haslett-Raftery both had low MSEs in all series greater than 500 data points in length. The step reductions in the MSE for the wavelet estimator can be clearly seen each time a new *octave* became available.

## 4. Application: Campito Mountain Data

The Campito Mountain bristlecone pine data is regarded as a standard example of a long memory time series. It is a 5405 year series of annual tree ring widths of bristlecone pines on Campito Mountain, California. It was studied by Baillie and Chung (2002) who determined that an ARFIMA(0,0.44,0) model fitted the data best. The lack of additional short term correlation in the data

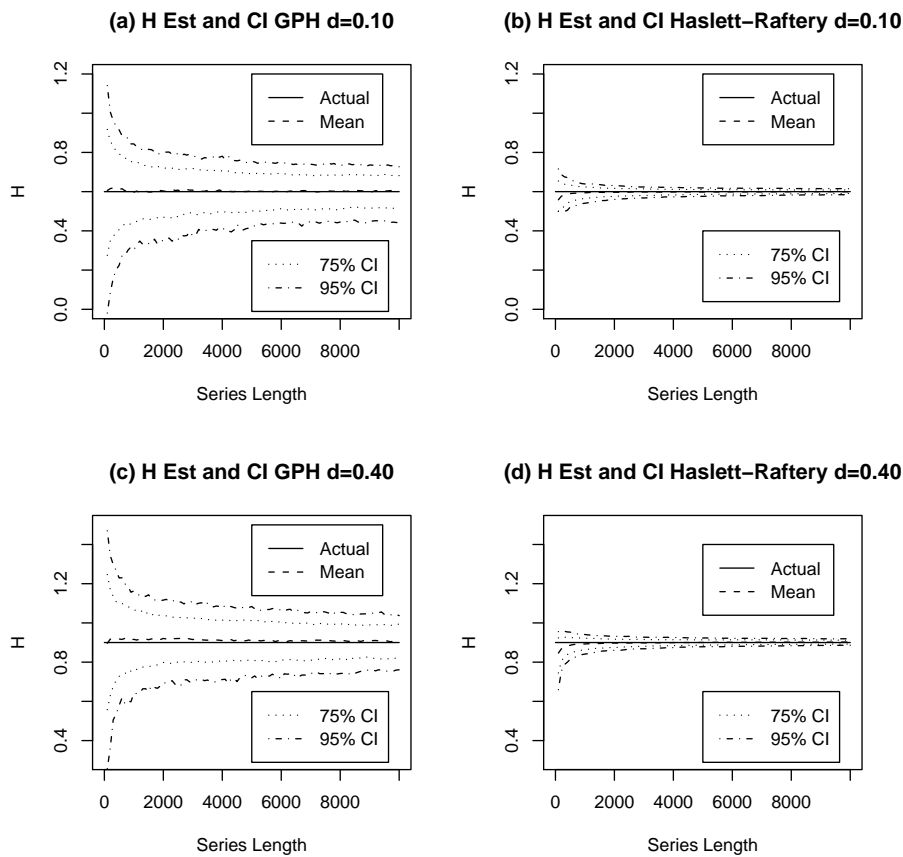

FIG 6. *Empirical confidence intervals for the H estimates with d = 0.10 (H = 0.60) and d = 0.40 (H = 0.90); (a) and (c) GPH estimator, (b) and (d) Haslett-Raftery estimator.*

means it is a good candidate for modeling with an FGN.

The Campito Mountain data is available in the R package tseries as the data set camp. We applied the 12 estimators to this series and estimated the goodness of fit to an FGN for all estimators where possible, except the Haslett-Raftery, using the test of Beran (1992) as implemented in the R package longmemo of Beran et al. (2006). The Beran test is more powerful against under estimation of $H$ than over estimation. The Beran test was unable to be used for $H$ values exceeding unity. It is important to note that Deo and Chen (2000) showed that the asymptotic properties of the test as presented by Beran (1992) were incorrect. We subjected the Beran test to a simulation study, the results of which will be presented at a later data. This study showed that for sample sizes of the order studied here the Beran test over rejects the null hypothesis by a small amount (e.g. typically 6 percent at the 5 percent level). Thus for pragmatic testing of goodness of fit, the Beran test can still be used with appropriate caution, alternatively critical values can be obtained through simulation.

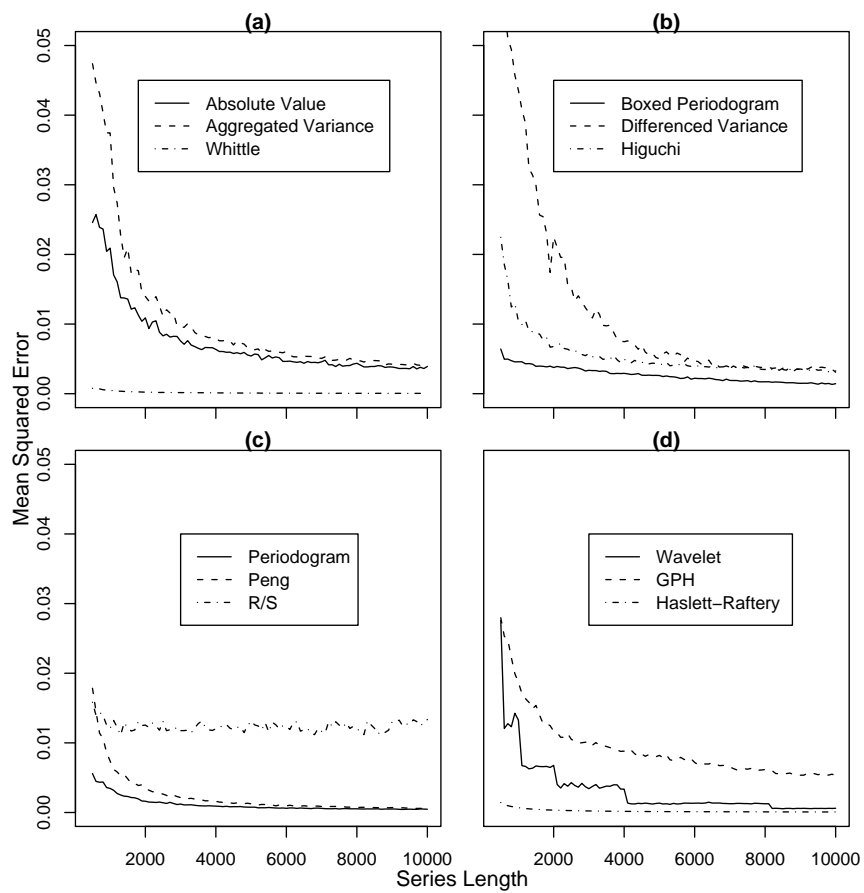

FIG 7. *Mean squared errors (MSE) as a function of series length for all 12 estimators with d=0.4 for the GPH and Haslett-Raftery and H=0.9 for the other ten. MSEs are reported starting at a series of 500 data points. (a) Absolute Value, Aggregated Variance and Whittle. (b) Boxed Periodogram, Difference Variance and Higuchi. (c) Peng, Periodogram and R/S. (d) Wavelet, GPH and Haslett-Raftery.*

| Method | H Est | Beran p-value | CPU Seconds | Expected H | Empirical p-value |
|---|---|---|---|---|---|
| Absolute Value | 0.862 | 0.435 | 0.19 | 0.831 | 0.70 |
| Aggregated Variance | 0.889 | 0.577 | 0.34 | 0.821 | 0.04* |
| Boxed Periodogram | 0.914 | 0.509 | 0.09 | 0.849 | 0.01** |
| Differenced Variance | 1.089 | - | 0.21 | 0.925 | 0.01** |
| GPH | 1.037 | - | 1.43 | 0.897 | 0.16 |
| Haslett-Raftery | 0.947 | 0.241 | 0.17 | - | - |
| Higuchi | 0.966 | 0.102 | 19.65 | 0.845 | $< 0.001$*** |
| Peng | 0.936 | 0.344 | 18.46 | 0.875 | $< 0.001$*** |
| Periodogram | 1.007 | - | 0.06 | 0.908 | $< 0.001$*** |
| Rescaled Range | 0.892 | 0.577 | 0.04 | 0.816 | 0.36 |
| Wavelet | 0.927 | 0.421 | 0.07 | 0.889 | 0.25 |
| Whittle | 0.876 | 0.540 | 1.05 | 0.890 | 0.15 |

TABLE 1
*The first two columns of results presents the H estimates and p-values returned by the Beran (1992) test for the Campito Mountain data for each of the 10 estimators of H and the GPH and Haslett-Raftery estimators of d converted to H equivalent. CPU times are in seconds on the SunBlade described in the text. The expected H column is the expected value that the estimator would report if the Campito data was an FGN with $H = 0.89$. The empirical p-value column is estimated empirically from the simulated data.*

The results are presented in Table (1). The Beran test indicated an $H$ value close to 0.89 fitted this data best. The maximum p-value was 0.577 for values of $H$ estimated by the aggregated variance and rescaled range estimators. Nine of the 12 estimators reported $H$ or $d$ values which lie in an acceptable range on the basis of the Beran (1992) test assuming we set our level of statistical significance at 0.05 to reject the null hypothesis of an FGN. The remaining three could not be tested.

Given the results from our simulated FGNs there were some unexpected $H$ estimates for the Campito data. On the basis of the simulations we expected the aggregated variance, absolute value, boxed periodogram, Higuchi, and rescaled range to return a low estimate for $H$. None of these estimators did so. As the Beran test reported that $H = 0.89$ yielded the best fit we used the median value from the simulations with series length 5400 and $H = 0.90$, and adjusted for the difference of 0.01 $H$ units, to estimate the value of $H$ which would be reported by each estimator if the data was from an FGN. This value is reported in Table (1) as "Expected H". The sixth column reports the empirically determined p-value for the actual estimate again using the simulated data. We do not report values for the Haslett-Raftery estimator as it estimates $d$ not $H$. It is interesting that six of the estimators reported $H$ estimates which are statistically significantly higher than their expected values.

The estimator which had the least bias and narrowest confidence interval in the simulations with series length 5400 and $H = 0.90$, namely the Whittle, was marginally out performed by the aggregated variance and rescaled range judged on the basis of the Beran test.

The fourth column of Table (1) reports CPU times in seconds. With present day computer speeds estimation times on the Campito series are not an issue. Only four estimators required more than one second of CPU time on this 5405 observation series. It is evident that some estimators which require longer compute times, such as the Higuchi and Peng, did not necessarily yield a more accurate estimation of $H$ for this data.

## 5. Discussion

It is clear from the simulations that not all estimators are created equal. Long memory occurs in the range $0.5 < H < 1$. Thus any estimator used to estimate the strength of the long memory needs to be both accurate and have a low variance.

The boxed periodogram method was developed specifically to deal with the problem of having most of the points used to estimate $H$ on the right-hand side of the graph. This was believed to, possibly, cause bias in the periodogram estimator. Beran (1994, p133) and Taqqu et al. (1995) outline some of the reasons such a method could be expected to be biased. In the series we investigated here the box periodogram estimator is inferior to the periodogram estimator it was intended to improve upon.

The differenced variance method was developed to be robust to trends which were known to cause spurious long memory in the R/S estimator (Bhattacharya et al., 1983). We did not test its robustness. Teverovsky and Taqqu (1999) established that the differenced variance method had a higher variance than the aggregated variance method, a result supported by our simulation study. In fairness to the method it must be pointed out that Teverovsky and Taqqu (1999) did not intend for it to be used alone but rather in conjunction with the aggregated variance method to test for the presence of shifting means or deterministic trends.

The Higuchi (1988) estimator only indirectly estimates $H$. It estimates the fractal dimension, $D$, of a series by estimating its path length. As implemented it then converts the estimate of $D$ to $H$ by the simple relationship $H = 2 - D$. This should be borne in mind if a researcher wishes to estimate $D$ rather than $H$ as it is a simple matter to recover $D$ from the $H$ estimate report by this implementation.

Taqqu et al. (1995) give a detailed proof that the method of Peng et al. (1994) is asymptotically unbiased. In the simulations the bias was never large but even at a sample size of 10,000 observations the estimator cannot be considered unbiased. However, its MSE approaches that of the Periodogram method as the series length increases which, in turn, is better than several others.

The wavelet estimator is asymptotically unbiased. In the simulations the bias was always small and was unbiased for series with longer than or equal to 4,100 observations.

## 6. Conclusions and Future Research

Of the twelve estimators examined here the Whittle and Haslett-Raftery estimators performed the best on simulated series. If we require an estimator to be close to unbiased across the full range of $H$ values for which long memory occurs and have a 95 percent confidence interval width of less than 0.1 $H$ or $d$ units (that is 20 percent of the range for $H$ or $d$ values in which long memory is observed), then for series with less than 4,000 data points they were the only two estimators worth considering. It should be noted that these estimators did not meet these criteria until the series lengths exceeded 700 and 1000 data points respectively. For series with 4,000 or more data points, the Peng estimator gave acceptable performance. For series with more than 7,000 data points the periodogram estimator was a worthwhile choice. For series with more than 8,200 data points the wavelet became a viable estimator. The remaining seven estimators did not give acceptable performance at any series lengths examined and are not recommended.

The Higuchi estimator is useful if the researcher wishes to recover the fractal dimension of the time series. In contrast to the other estimators it provides useful information about a time series if the series is not an FGN (or FI(d)) series. As an estimator of $H$ it is inferior to several others.

The boxed periodogram method is clearly inferior to the periodogram method it was intended to improve upon for FGNs. Further research would be needed to test if it is more robust than the periodogram method in series with departures from a pure FGN. This could be accomplished, for example, by simulating ARFIMA series with non-zero AR and MA components or series with structural breaks.

The R/S estimator is of considerable historical interest but had a major deficiency in that its MSE plateaued while all other estimators' MSEs decreased with increasing series length. Against this we must note that it was one of the two best performing estimators when applied to the Campito data when judged by the Beran test.

The differenced variance estimator was the worst of the twelve estimators in short series. For series longer than 6,000 data points its MSE was better than the R/S and on a par with the absolute value, aggregated variance and Higuchi methods. As noted above, Teverovsky and Taqqu (1999) do not recommend its use in isolation as it is part of a test for shifting means or deterministic trends. Teverovsky and Taqqu (1999) also recommend the aggregated and differenced variance plots always be examined visually. We agree with these recommendations. We did not test its robustness to shifting means or deterministic trends. Some numerical results on its performance in these two situations can be found in Teverovsky and Taqqu (1999).

The application to the Campito data six of the estimators reported statistically significantly different $H$ estimates than expected based on the evidence from the simulated series. Although the fit of the Campito data to an FGN is good (p=0.577), these six estimators do not seem to be robust to whatever specific departures from an FGN that are present in the data. This suggests that a researcher should not rely on a single estimator when estimating $H$ and that the Beran (1992) test should always be applied to test the goodness of fit of the data to an FGN, while being aware that its asymptotic properties are currently unknown.

Because of the apparent lack of robustness, a useful avenue of future research would be to quantify the sensitivity of these estimators to various types of departures from an FGN, e.g. FGN series with a small number of shifts in mean or a small number of outlier data points.